\documentclass[aps,preprint,nofootinbib,superscriptaddress]{revtex4}
\usepackage{amssymb}

\usepackage{epsfig}
\usepackage[bookmarksnumbered,bookmarksopen,colorlinks,citecolor=blue,linkcolor=blue]{hyperref}
\begin{document}


\title{Mirror nuclei constraint in mass formula }

\author{Ning Wang}
\email{wangning@gxnu.edu.cn}  \affiliation{Department of Physics,
Guangxi Normal University, Guilin 541004, P. R. China}

\author{Zuoying Liang}
\affiliation{Department of Physics, Guangxi Normal University,
Guilin 541004, P. R. China}

\author{Min Liu}
\affiliation{Department of Physics, Guangxi Normal University,
Guilin 541004, P. R. China}
\affiliation{ College of Nuclear
Science and Technology, Beijing Normal University, Beijing,
100875, P. R. China}

\author{Xizhen Wu}
\affiliation{China Institute of Atomic Energy, Beijing 102413, P.
R. China}

\begin{abstract}
The macroscopic-microscopic mass formula is further improved by
considering mirror nuclei constraint. The rms deviation with
respect to 2149 measured nuclear masses is reduced to 0.441 MeV.
The shell corrections, the deformations of nuclei, the neutron and
proton drip lines, and the shell gaps are also investigated to
test the model. The rms deviation of $\alpha$-decay energies of 46
super-heavy nuclei is reduced to 0.263 MeV. The central position
of the super-heavy island could lie around $N=176 \sim 178$ and
$Z=116\sim 120$ according to the shell corrections of nuclei.
\end{abstract}

\maketitle

\begin{center}
\textbf{I. INTRODUCTION}
\end{center}

The concept of symmetry in physics is a very powerful tool for
understanding the behavior of Nature. The isospin symmetry
discovered by Heisenberg plays an important role in nuclear
physics. In the absence of Coulomb interactions between the
protons, a perfectly charge-symmetric and charge-independent
nuclear force would result in the binding energies of mirror
nuclei (i.e. nuclei with the same atomic number $A$ but with the
proton number $Z$ and neutron number $N$ interchanged) being
identical \cite{Lenzi,Shlomo}. Although the Coulomb interaction
can result in the isospin-symmetry-breaking (ISB), the measured
energy differences in the excited analogue states between mirror
nuclei (MED) amount to tens of keV and do not generally exceed 100
keV, which indicates that the "nuclear part" of the binding
energies in pairs of mirror nuclei should be close to each other,
i.e.
\begin{equation}
E_B-E_C \approx E_B^{\prime }-E_C^{\prime}.
\end{equation}
Where, $E_B$ and $E_C$ denote the total energy and the Coulomb
energy of a nucleus, respectively, and $E_B^{\prime }$ and
$E_C^{\prime}$ denote the corresponding values of the mirror
nucleus. Combining the macroscopic-microscopic mass formula and
Eq.(1), one can obtain the constraint between the shell
corrections of the mirror nuclei,
\begin{equation}
|\Delta E -\Delta E^{\prime}| \approx 0,
\end{equation}
that is to say, a small value for the difference of the shell
corrections of a nucleus and of its mirror nucleus. It is
interesting to study the constraint between mirror nuclei and the
ISB effect for improving the nuclear mass formula, especially for
the calculations of neutron-rich nuclei and super-heavy nuclei.

In addition, the influence of the Coulomb interaction on the
single-particle levels attracted a lot attention in recent years.
It has been shown that single-particle effects, induced by the
electromagnetic spin-orbit interaction and the Coulomb orbital
term, produce large effects in the MED for nuclei in the upper
\emph{sd} and \emph{fp} shells \cite{Lenzi06}. In \cite{Cwiok},
the authors found that the Coulomb potential strength does not
change the position of magic gaps 50, 82 and 126, but strongly
influences the shell structure of super-heavy nuclei. These
investigations show that it is necessary to study the influence of
the Coulomb term on the isospin-symmetry-breaking and on the
binding energies of nuclei. The aim of the present work is to
improve the semi-empirical mass formula through studying the
mirror nuclei constraint due to the isospin-symmetry and the
influence of the Coulomb term on the single-particle levels and
consequently on the shell corrections of nuclei. The paper is
organized as follows: In Sec.II, we introduce the semi-empirical
nuclear mass formula and some modifications in this work. In Sec.
III, some results with the proposed model are presented. Finally,
conclusions and discussions are contained in Sec.IV.

\begin{center}
\textbf{II. MODIFICATIONS OF THE MASS FORMULA}
\end{center}

In \cite{Wang}, we proposed an semi-empirical nuclear mass formula
based on the macroscopic-microscopic method \cite{Moll95}. The total
energy of a nucleus can be calculated as a sum of the liquid-drop
energy and the Strutinsky shell correction $\Delta E$,
\begin{eqnarray}
E (A,Z,\beta)=E_{\rm LD}(A,Z) \prod_{k \ge 2} \left (1+b_k
\beta_k^2 \right )+\Delta E (A,Z,\beta).
\end{eqnarray}
The liquid drop energy of a spherical nucleus $E_{\rm LD}(A,Z)$ is
described by a modified Bethe-Weizs\"acker mass formula,
\begin{eqnarray}
E_{\rm LD}(A,Z)=a_{v} A + a_{s} A^{2/3}+ E_C + a_{\rm sym} I^2 A +
a_{\rm pair}  A^{-1/3}\delta_{np}
\end{eqnarray}
with isospin asymmetry $I=(N-Z)/A$, and the symmetry energy
coefficient,
\begin{eqnarray}
 a_{\rm sym}=c_{\rm sym}\left [1-\frac{\kappa}{A^{1/3}}+\frac{2-|I|}{ 2+|I|A} \ \right
 ].
\end{eqnarray}
The isospin dependence of the pairing term is also considered (see
the expression of $\delta_{np}$ in \cite{Wang} for details). The
terms with $b_k$ describe the contribution of nuclear deformation to
the macroscopic energy, and the mass dependence of $b_k$ is written
as,
\begin{eqnarray}
b_k=\left ( \frac{k}{2} \right ) g_1A^{1/3}+\left ( \frac{k}{2}
\right )^2 g_2 A^{-1/3}.
\end{eqnarray}
The shell correction is obtained by the traditional Strutinsky
procedure \cite{Strut} by setting the order $p=6$ of the
Gauss-Hermite polynomials and the smoothing parameter
$\gamma=1.2\hbar\omega_0$ with $\hbar\omega_0=41 A^{-1/3}$ MeV. For
the deformation of nuclei, we only consider axially-deformed cases.

In this work, we make the following modifications to the mass
formula:
\begin{itemize}
\item  The Coulomb energy form is slightly changed, $Z(Z-1)$ is
replaced by $Z^2$,
\begin{eqnarray}
E_C=a_{c}\frac{Z^{2}}{A^{1/3}} \left [ 1- Z^{-2/3} \right],
\end{eqnarray}
following the form in the finite range droplet model (FRDM)
\cite{Moll95}. This modification can slightly improve the rms
deviation with respect to 2149 measured nuclear masses \cite{Audi}
of nuclei [N and $Z \ge 8$] by about $2\sim 3\%$.

\item  The microscopic shell correction of a nucleus is modified
as,
\begin{eqnarray}
\Delta E=c_1 E_{\rm sh} + |I| E_{\rm sh}^{\prime}.
\end{eqnarray}
Where, $E_{\rm sh}$ and $E_{\rm sh}^{\prime}$ denote the shell
energy of a nucleus and of its mirror nucleus, respectively. The
additionally introduced $|I| E_{\rm sh}^{\prime}$ term is to
empirically take into account the mirror nuclei constraint and the
isospin-symmetry-breaking effect. We find that this term can
considerably reduce the rms deviation of masses by about $10\%$. The
isospin-dependence in Eq.(8) is to consider the increase of the
difference between neutron-neutron and proton-proton pairs in
neutron-rich or proton-rich nuclei. The $|I|E_{\rm sh}^{\prime}$
term can effectively reduce the shell correction deviation $|\Delta
E -\Delta E^{\prime}|$ in pairs of mirror nuclei, which is required
from the constraint in Eq.(2) and is helpful to restore the isospin
symmetry in the mirror nuclei.  If without the $|I|E_{\rm
sh}^{\prime}$ term in Eq.(8), we obtain
\begin{eqnarray}
|\Delta E-\Delta E^{\prime}|= c_1 |E_{\rm sh} - E_{\rm
sh}^{\prime}|.
\end{eqnarray}
Considering the $|I|E_{\rm sh}^{\prime}$ term in $\Delta E$, we
obtain
\begin{eqnarray}
|\Delta E-\Delta E^{\prime}|& = &(c_1 - |I|) |E_{\rm sh} - E_{\rm
sh}^{\prime}| \nonumber \\
&\leq& c_1 |E_{\rm sh} - E_{\rm sh}^{\prime}|.
\end{eqnarray}

\begin{figure}
\includegraphics[angle=-0,width= 0.75\textwidth]{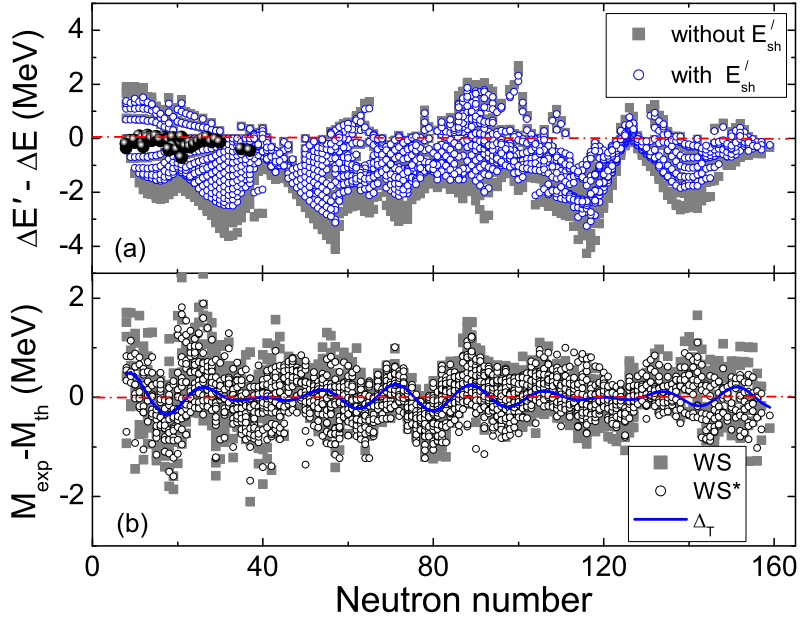}
 \caption{(Color online) (a) Shell correction difference $\Delta E^{\prime}
-\Delta E $ in pairs of mirror nuclei. The squares and the open
circles denote the results without and with the $E_{\rm
sh}^{\prime}$ term being taken into account, respectively. The balls
denote the experimental values of $(E_B-E_C)-(E_B^{\prime
}-E_C^{\prime})$ between mirror nuclei with the Coulomb energy form
in \cite{Wang}. (b) Deviations of the calculated nuclear masses from
the experimental data \cite{Audi}. The squares and the crosses
denote the results of WS and WS*, respectively. The solid curve
denote the results of an empirical formula, $\Delta_T=-0.7 \left
[\cos\left (2\pi \frac{N}{16}\right )+\cos \left (2 \pi \frac{N}{20}
\right ) \right] A^{-1/3}$, which will be discussed later.}
\end{figure}
To illustrate this point, in Fig.1(a) we show the values of $ \Delta
E^{\prime} -\Delta E$ between mirror nuclei as a function of neutron
number with the WS model \cite{Wang}. The balls denote the
experimental values of the nuclear energy difference
$(E_B-E_C)-(E_B^{\prime }-E_C^{\prime})$ between mirror nuclei by
adopting the Coulomb energy form in the WS model. One can see that
the experimental binding energies of  pairs of mirror nuclei are
indeed close to each other as mentioned previously when removing the
Coulomb energies and the deviations are generally smaller than 1
MeV. The squares and the circles denote the results of shell
correction difference without and with the $|I|E_{\rm sh}^{\prime}$
term being taken into account, respectively. Here, the shell energy
of a nucleus is calculated at the deformation of its mirror nucleus
for the sake of simplicity, since the deformations of pairs of
mirror nuclei are close to each other for most nuclei according to
the calculated results with WS. The WS calculations show that the
shell correction differences caused by the Coulomb potentials are
larger than 3 MeV for some mirror nuclei, which obviously
over-predict the experimental nuclear energy differences [see the
balls in Fig.1(a)]. The electromagnetic spin-orbit interaction, the
Coulomb orbital term or the Coulomb potential strength are therefore
introduced by some authors \cite{Lenzi06,Cwiok} for improving the
traditional Coulomb potential as mentioned previously. In this work,
the influence of the Coulomb term is effectively considered by
introducing the shell energy of the mirror nuclei. The shell
correction difference between mirror nuclei is effectively reduced
by about 1 MeV after the $|I|E_{\rm sh}^{\prime}$ term being
considered in $\Delta E$.

\item  The $\beta_6$ deformation of nuclei is taken into account,
which slightly improves the results of heavy nuclei.
 \end{itemize}

\begin{center}
\textbf{III. RESULTS}
\end{center}

With these modifications and the obtained optimal parameters of mass
formula which are listed in Table 1 and labelled as WS*, the rms
deviations of the 2149 nuclear masses is further reduced by $15\%$,
to 0.441 MeV and the rms deviations of the neutron separation
energies of 1988 nuclei is reduced to 0.332 MeV (see Table 2). Fig.1
(b) shows the deviations of the calculated nuclear masses from the
experimental data. Considering the shell constraint between mirror
nuclei (WS*), the results are effectively improved.

\begin{table}
\caption{ Model parameters of the mass formula. }
\begin{tabular}{ccc }
\hline\hline
  parameter                        & ~~~~~~~ WS~~~~~~~~  & ~~~WS*~~~\\ \hline
 $a_v  \; $ (MeV)                  &  $-15.5841$ &   $-15.6223$   \\
 $a_s \; $  (MeV)                  &   18.2359   &   18.0571 \\
 $a_c \; $ (MeV)                   &   0.7173    &   0.7194 \\
 $c_{\rm sym} $(MeV)               &   29.2876   &   29.1563   \\
 $\kappa \;  $                     &   1.4492    &   1.3484  \\
 $a_{\rm pair} $(MeV)              &   $-5.5108$ &   $-5.4423$   \\
 $g_1 $                            &   0.00862   &   0.00895 \\
 $g_2 $                            &  $-0.4730$  &   $-0.4632$   \\
 $c_1  \; $                        &  0.7274     &   0.6297   \\
 $V_0$ (MeV)                       &  $-47.4784$ &   $-46.8784$ \\
 $r_0$ (fm)                        &  1.3840     &   1.3840   \\
 $a $ (fm)                         &  0.7842     &   0.7842  \\
 $\lambda_0$                       &  26.3163    &   26.3163 \\

 \hline\hline
\end{tabular}
\end{table}

\begin{table}
\caption{ rms $\sigma$ deviations between data AME2003 \cite{Audi}
and predictions of several models (in MeV). The line $\sigma (M)$
refers to all the 2149 measured masses, the line $\sigma (S_n)$ to
the 1988 measured neutron separation energies $S_n$. The calculated
masses with FRDM are taken from \cite{Moll95}. The masses with
HFB-14 and HFB-17 are taken from \cite{HFB14} and \cite{HFB17},
respectively. WS*+$\Delta_T$ means the correction $\Delta_T$ for
empirically considering the tetrahedral deformation is added to the
binding energy of a nucleus with WS*.}
\begin{tabular}{ccccccc}
 \hline\hline
     & ~FRDM~ & HFB-14 & HFB-17 & ~~WS~~ & ~~WS*~~& ~~WS*+$\Delta_T$~~\\
\hline
 $\sigma  (M)$      & $0.656$ & $0.729$ & $0.581$  & $0.516 $ & $0.441 $ & $0.417 $\\
 $\sigma  (S_n)$    & $0.399$ & $0.598$ & $0.506$  & $0.346 $ & $0.332 $ & $0.330 $\\

 \hline\hline
\end{tabular}
\end{table}

\begin{figure}
\includegraphics[angle=-0,width= 1\textwidth]{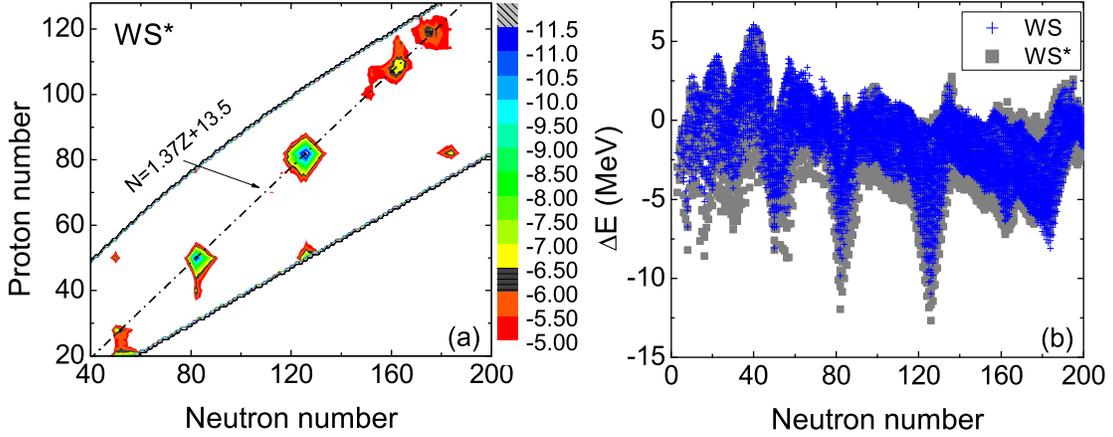}
 \caption{(Color online) (a) Contour plot of shell corrections of nuclei with
 WS*. The dot-dashed line passes through the areas with the known heavy magic nuclei.
 (b) Shell corrections of nuclei as a function of neutron number.
 The crosses and the squares denote the results of WS and WS*, respectively.}
\end{figure}

\begin{figure}
\includegraphics[angle=-0,width= 1\textwidth]{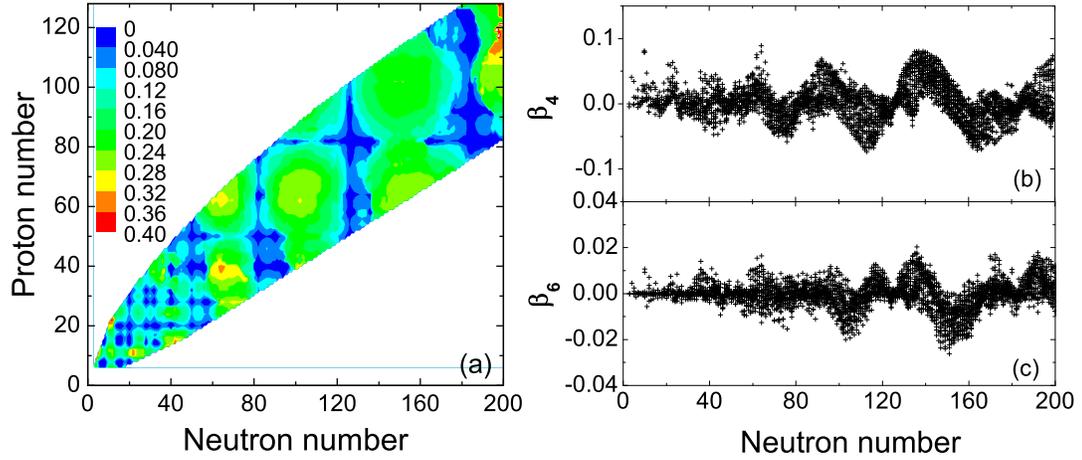}
 \caption{(Color online) (a) Contour plot of quadrupole deformation  $|\beta_2|$  of nuclei with WS*.
 (b)  $\beta_4$ and (c) $\beta_6$ of nuclei as a function of neutron number, respectively.}
\end{figure}

\begin{figure}
\includegraphics[angle=-0,width= 0.7\textwidth]{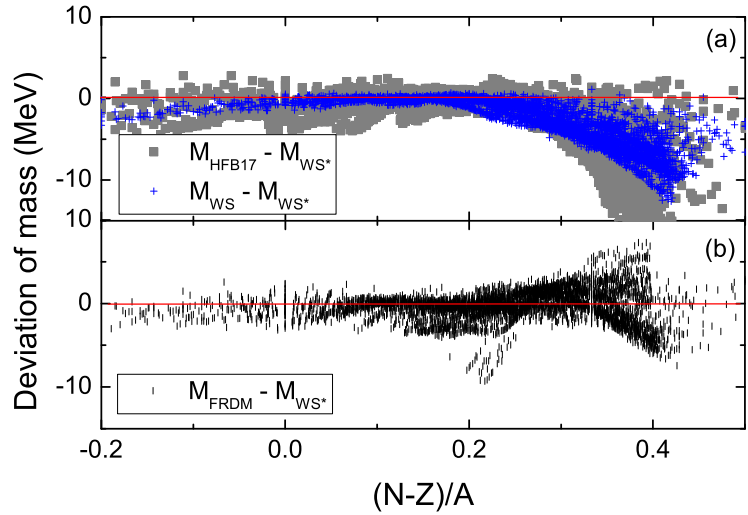}
 \caption{(Color online) Deviations of calculated nuclear masses in this work from the results of other models.  The
calculated masses with FRDM and HFB-17 are taken from \cite{Moll95}
and \cite{HFB17}, respectively.}
\end{figure}

\begin{figure}
\includegraphics[angle=-0,width= 0.7\textwidth]{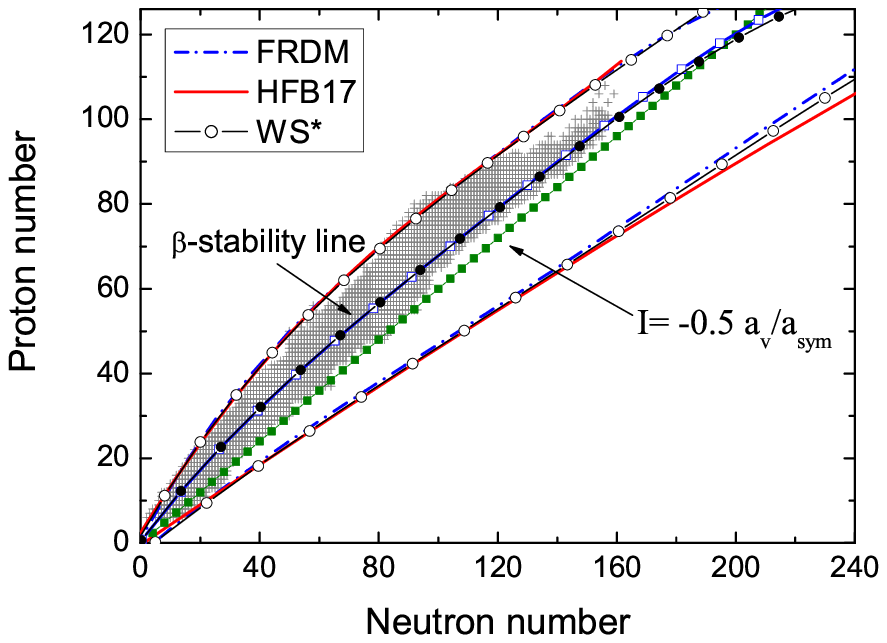}
 \caption{(Color online) Smooth drip lines from different mass formulas. The crosses
denote the measured nuclei. The solid-squared line denote the
neutron drip line of nuclei $A \rightarrow \infty $ with
$I=-\frac{1}{2}\frac{a_v}{a_{\rm sym}}$ and taking $a_v=-16$ MeV
and $a_{\rm sym}=32$ MeV. The open-squared and the solid-circled
curve denote the smooth $\beta$-stability line from FRDM and WS*
calculations, respectively. }
\end{figure}

In Fig.2 we show the calculated shell corrections $\Delta E$ of
nuclei with this model. Considering the shell constraint between
mirror nuclei, the nuclei with the largest shell corrections in
the super-heavy region slightly moves to $N=176$ and $Z=120$. The
shell energies with WS* for nuclei around $N=16$ and $N=28$ become
larger in absolute values, whilst those for nuclei around
($N=184$, $Z=82$) and $^{100}$Sn become smaller compared with the
WS calculations. In Fig.3, we show the calculated deformations of
nuclei with WS*. Obviously, the calculated structure of the known
magic nuclei is spherical in shape. For light nuclei, the
$\beta_6$ deformations of nuclei are not very obvious, compared
with the intermediate and heavy nuclei. In Fig.4, we show the
deviations of the calculated nuclear masses in this work from the
results of other models as a function of isospin asymmetry. One
can see that for highly neutron-rich nuclei ($I>0.3$) the
deviations from these different models are large, and the results
from FRDM and WS* are relatively close to each other, while the
results from WS are relatively close to those of HFB-17
\cite{HFB17}.

In Fig.5, we show the drip lines obtained with different mass
formulas. To remove the fluctuations due to the shell and pairing
effects, we do a polynomial fitting to the calculated results with
the FRDM, the HFB-17, and the WS* models, respectively. The
leftmost and rightmost curves denote the smooth proton drip line
 (for odd-Z nuclei) and the smooth neutron drip line, respectively.
The crosses denote the measured nuclei. For the proton drip line,
the three models give similar results. For the neutron drip line,
the results slightly deviate from each other at heavy mass region.
Based on the liquid-drop model, the neutron separation energy of
an intermediate and heavy nucleus ($A\gg1$) can be approximately
written as,
\begin{eqnarray}
S_n \simeq -a_v-2 a_{\rm sym} I.
\end{eqnarray}
Where, $a_v$ (negative value) and $a_{\rm sym}$ (positive value) are
the coefficients of the volume energy  and the symmetry energy of a
nucleus, respectively. For the neutron drip line ($S_n=0$) of
intermediate and heavy mass region, we obtain the isospin asymmetry
at the drip line
\begin{eqnarray}
 I_{nd} \simeq -\frac{1}{2}\frac{a_v}{a_{\rm sym}}.
\end{eqnarray}
One can see that the neutron drip line directly relates to the ratio
of $a_v$ to $a_{\rm sym}$. The difference of the neutron drip line
from different models is probably due to the difference of the
coefficients $a_v$ and $a_{\rm sym}$ adopted in the models. For
nuclei with $A \rightarrow \infty $ or asymmetric nuclear matter
($a_v\approx -16$ MeV and $a_{\rm sym} \approx 32$ MeV), we obtain
the corresponding neutron drip line which is also shown in Fig.5
(solid-squared line). From the figure, one can see that most of
measured nuclei are located in the left side of the solid-squared
line. In addition, we show the smooth $\beta$-stability line from
FRDM (open-squared curve) and WS* (solid-circled curve)
calculations, respectively. At $Z=120$, the corresponding neutron
number of the nuclei along the $\beta$-stability line is about
$N=200$.

\begin{figure}
\includegraphics[angle=-0,width= 0.85\textwidth]{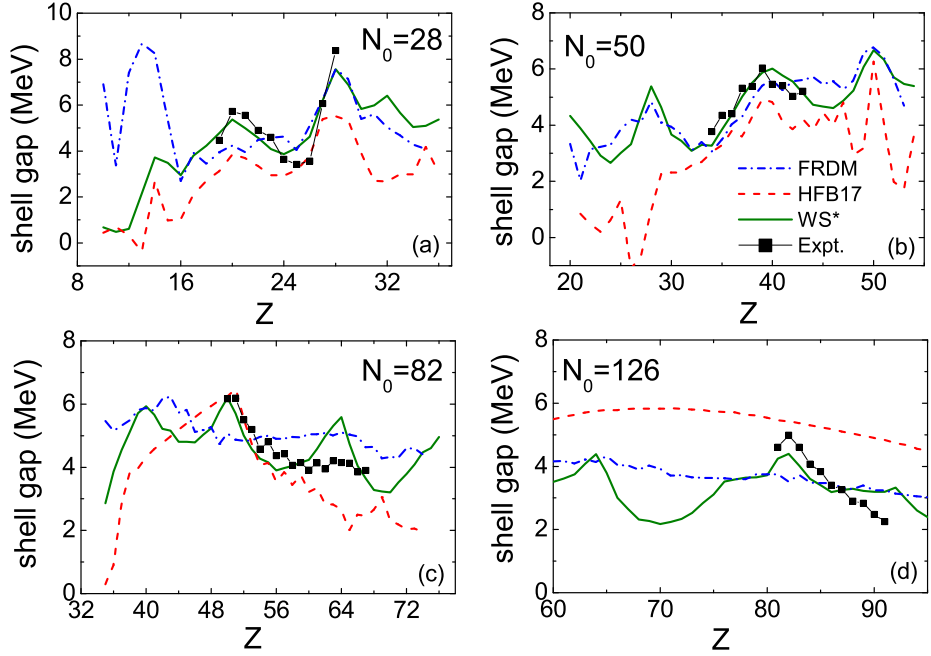}
 \caption{(Color online) Shell gap calculated with different models. The dashed,
 the  dot-dashed and the solid curve denote the results of HFB-17, FRDM and WS*,
 respectively. The squared curve denote the experimental data.  }
\end{figure}

To further test the model, we study the shell gaps. As a measure of
the discontinuity in the two neutron separation energy $S_{2n}$ at
magic neutron numbers $N_0$, the shell gap \cite{Lun03},
\begin{eqnarray}
 \Delta_n(N_0,Z)=S_{2n}(N_0,Z)-S_{2n}(N_0+2,Z),
\end{eqnarray}
is a sensitive quantity to test the model. In Fig.6, we show the
calculated shell gaps at the magic neutron numbers
$N_0=28,50,82,126$
 with different models. The dashed, the dot-dashed and the
solid curve denote the results of HFB-17, FRDM and WS*,
respectively. The squared curve denote the experimental data. The
most shell gaps can be reasonable well described by the WS* model,
except the shell gap at sub-shell closure $Z=64$ which is
over-predicted by WS* and FRDM and is under-predicted by the HFB-17
model. In Fig.6(b), the peak (large shell gap) at magic number
$Z=28$ disappears according to the HFB-17 calculations, and the peak
at $Z=82$ can not be reasonably well described  from the FRDM and
HFB-17 calculations in Fig.6(d). The experimental shell gaps at
magic numbers $Z=20$, 28, 40, 50, 82 can be remarkably well
described with the proposed model.

\begin{figure}
\includegraphics[angle=-0,width= 0.9\textwidth]{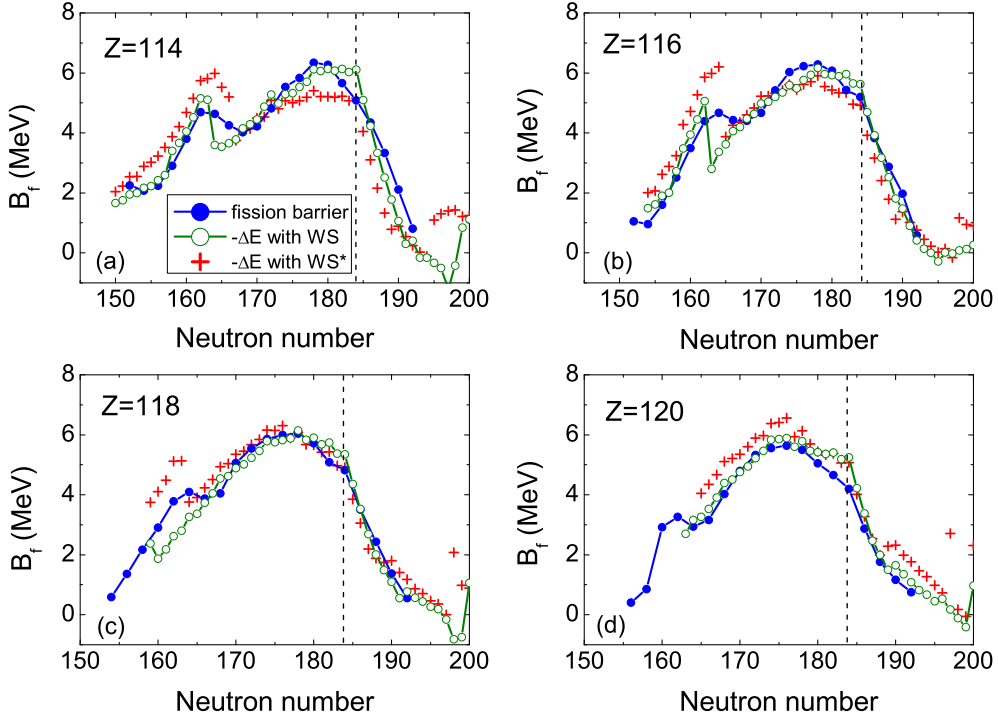}
 \caption{(Color online) Fission barriers of some super-heavy nuclei.
 The solid-circled curve denote the calculated fission barriers of super-heavy nuclei in Ref.\cite{Kowal}.
 The open-circled curve and the crosses denote the values of $-\Delta E$ with WS and WS*, respectively.
 The dashed line denotes the position $N=184$.}
\end{figure}

In addition, we study the relation between the fission barrier of
super-heavy nucleus and the corresponding shell correction of the
nucleus. Neglecting the shell energy at the saddle point, the
fission barrier of a nucleus can be approximately written as
\cite{Hivap},
\begin{eqnarray}
 B_f \approx B_{\rm LD}-\Delta E.
\end{eqnarray}
Where $B_{\rm LD}$ and $\Delta E$ are the macroscopic fission
barrier and the shell correction of a nucleus at its ground state.
For super-heavy nucleus, the macroscopic fission barrier generally
disappears and consequently the fission barrier can be roughly
evaluated through the corresponding shell correction of the
nucleus. In Fig.7, we show the fission barriers of a number of
super-heavy nuclei (solid-circled curve) \cite{Kowal} which are
calculated with the macroscopic-microscopic approach, considering
the deformation of system up to $\beta_8$ and the triaxial
deformation. The open-circled curve and the crosses denote the
values of $-\Delta E$ with WS and WS*, respectively. One can see
that the calculated fission barriers are generally close to the
values of $-\Delta E$ with WS. Both of models in which the mirror
nuclei constraint is not taken into account, predict the two
neutron magic numbers $N=162$ and $N=178$ at the super-heavy
region. When the constraint between mirror nuclei is considered
(WS*), the results for nuclei with $Z=116$ and 118 do not change
too much around $N=178$, while the results for nuclei with $Z=114$
and 120 change about 1 MeV. These investigates indicate: 1) The
calculated shell corrections (in absolute value) with the proposed
mass formula  are comparable to the fission barriers of
super-heavy nuclei; 2) The mirror nuclei constraint could
influence the shell structure of nuclei with $Z=114$ and 120. In
Fig.7, the dashed lines denotes the position $N=184$. For nuclei
with $N>184$, the fission barriers fall rapidly with the increase
of neutron number. In addition, the shell corrections (in absolute
value) of nuclei with $Z>120$ are obviously smaller than that of
$^{292}120$. According to the calculated shell corrections of
nuclei, the central position of the super-heavy island could lie
around $N=176\sim 178$ and $Z=116\sim 120$.

Furthermore, we study the $\alpha$-decay energies of 46
super-heavy nuclei (the experimental data are taken from Ref.
\cite{Ogan}, Table I of Ref.\cite{Dong} and Table II of
Ref.\cite{Ma}). The rms deviation of the $\alpha$-decay energies
of the 46 super-heavy nuclei falls from 0.566 MeV with FRDM to
0.263 MeV with the WS* model (the corresponding result with WS is
0.284 MeV). In Table III, we list the $\alpha$-decay energies
$Q_{\alpha}$ and the shell corrections $\Delta E$ in 6
$\alpha$-decay chains of super-heavy nuclei with $Z=117$
\cite{Ogan} and $Z=120$. The available experimental data can be
reproduced reasonably well. These calculations indicate that the
proposed mass formula is relatively reliable for description of
the masses of super-heavy nuclei.

\begin{table}
\caption{ $\alpha$-decay energies $Q_{\alpha}$ and the shell
corrections in 6 $\alpha$-decay chains with WS* (in MeV). The
experimental data are taken from \cite{Ogan,Dong}.}
\begin{tabular}{cccccccccc }
 \hline\hline

 $~~A~~$ & ~~$Z$~~ &  $Q_{\alpha}$(Expt.) & $Q_{\alpha}$(WS*) & $ \Delta E $(WS*) &~~ $A$ ~~& ~~$Z$ ~~&  $Q_{\alpha}$(Expt.) & $Q_{\alpha}$(WS*) & $ \Delta E $(WS*) \\
\hline
 294  & 117  &  10.96(10) & 11.32 & -5.77 &  293  & 117  &  11.18(8) & 11.62  & -5.70 \\
 290  & 115  &  10.09(40) & 10.38 & -5.17 &  289  & 115  &  10.45(9) & 10.34  & -5.28 \\
 286  & 113  &   9.77(10) &  9.82 & -4.33 &  285  & 113  &   9.88(9) & 10.13  & -4.46 \\
 282  & 111  &   9.13(10) &  9.62 & -4.10 &  281  & 111  &         - & 10.04  & -4.17 \\
 278  & 109  &   9.69(19) &  9.66 & -4.88 &  277  & 109  &         - &  9.75  & -5.07 \\
 274  & 107  &   8.93(10) &  8.71 & -4.80 &  273  & 107  &         - &  8.98  & -5.05 \\
\hline
 296  & 120  &          - &  13.25 & -6.56 &  298  & 120  &         - & 12.81 & -6.13 \\
 292  & 118  &          - &  12.06 & -6.16 &  294  & 118  &  11.81(6) & 12.15 & -6.31 \\
 288  & 116  &          - &  11.31 & -5.32 &  290  & 116  &  11.00(8) & 11.12 & -5.61 \\
 284  & 114  &          - &  10.93 & -4.50 &  286  & 114  &  10.33(6) & 10.25 & -5.09 \\
 280  & 112  &          - &  10.90 & -4.51 &  282  & 112  &         - & 10.27 & -4.68 \\
\hline
 304  & 120  &          - &  12.49 & -5.08 &  320  & 120  &         - &  9.85 & -2.31 \\
 300  & 118  &          - &  11.70 & -5.43 &  316  & 118  &         - &  9.27 & -2.08 \\
 296  & 116  &          - &  10.98 & -5.43 &  312  & 116  &         - &  8.73 & -0.14 \\
 292  & 114  &          - &   9.12 & -5.40 &  298  & 114  &         - &  8.07 & -5.15 \\
 288  & 112  &          - &   9.36 & -4.46 &  294  & 112  &         - &  8.38 & -3.77 \\

 \hline\hline
\end{tabular}
\end{table}

\begin{center}
\textbf{IV. CONCLUSION AND DISCUSSION}
\end{center}

In summary, the semi-empirical mass formula based on the
macroscopic-microscopic method has been further improved by
considering the constraint between mirror nuclei. The rms
deviation with respect to 2149 measured nuclear masses is reduced
to 0.441 MeV and the rms deviation of the neutron separation
energies of 1988 nuclei falls to 0.332 MeV. The shell corrections,
the deformations of nuclei and the neutron and proton drip lines
have been investigated also. The predicted central position of the
super-heavy island according to the calculated shell corrections
of nuclei could lie around $N=176\sim 178$ and $Z=116\sim 120$,
considering the mirror nuclei constraint. The shell corrections of
super-heavy nuclei (in absolute value) are close to the
corresponding fission barriers of the nuclei from other
macroscopic-microscopic model. The shell gaps at proton magic
numbers $Z=20$, 28, 40, 50, 82 can be remarkably well described
with the proposed model. The rms deviation of the $\alpha$-decay
energies of 46 super-heavy nuclei is reduced from 0.566 MeV with
FRDM to 0.263 MeV with the proposed model in this work.

In addition, we note that the deviations from the measured masses
for some nuclei with $N \approx 18$,26,40,56,64,70,80,88 etc. are
relatively large, with both WS and WS*, which may be caused by the
triaxial deformation of nuclei or the tetrahedral symmetry in
nuclei \cite{Cur,Dudek}. It is found that the strongest
tetrahedral-symmetry effects appear at tetrahedral-magic numbers
16, 20, 32, 40, 56, etc., and the tetrahedral deformation can
bring over a few MeV of energy gain in the nucleus \cite{Dudek}.
We empirically describe the influence of the tetrahedral
deformation on the binding energies of nuclei by using two cosine
functions together with the two tetrahedral-magic numbers 16 and
20, $ \Delta_T=-0.7 \left [\cos\left (2\pi \frac{N}{16}\right
)+\cos \left (2 \pi \frac{N}{20} \right ) \right] A^{-1/3}$ MeV.
The solid curve in Fig.1 (b) denotes the results of $\Delta_T$.
With the empirical function $\Delta_T$, the rms deviation of 2149
nuclear masses can be further reduced by $5\%$, to 0.417 MeV.
Microscopic study on the triaxial deformation of nuclei is in
progress.

\newpage

\begin{center}
\textbf{ACKNOWLEDGEMENTS}
\end{center}

We thank Prof. Zhuxia Li and Prof. Xiaohong Zhou for valuable
suggestions. This work was supported by National Natural Science
Foundation of China, Nos 10875031, 10847004 and 10865002. The
obtained mass table with the proposed formula is available from
http://www.imqmd.com/wangning/WS3.3.zip.


\begin{thebibliography}{99}

\bibitem{Lenzi} S. M. Lenzi  and M. A. Bentley, Lecture Notes in
Physics, \textbf{764}, 57 (2009).

\bibitem{Shlomo} S. Shlomo, Rep. Prog. Phys. \textbf{41}, 957 (1978)

\bibitem{Lenzi06} S. M. Lenzi, J. Phys.: Conf. Seri. \textbf{49},
85 (2006).

\bibitem{Cwiok} S. $\acute{\rm C}$wiok, J. Dobaczewski, et al., Nucl. Phys. A
\textbf{611}, 211 (1996).


\bibitem{Wang} Ning Wang, Min Liu and Xizhen Wu, Phys. Rev. C
\textbf{81}, 044322 (2010).

\bibitem{Moll95} P. M\"oller, J. R. Nix, et al., At. Data and
Nucl. Data Tables \textbf{59}, 185  (1995).


\bibitem{Strut} V. M. Strutinsky  and F. A. Ivanjuk, Nucl. Phys. A
\textbf{255}, 405 (1975).

\bibitem{Audi} G. Audi, A.H. Wapstra and C. Thibault, Nucl. Phys. A
\textbf{729}, 337 (2003).

\bibitem{HFB14} S. Goriely, M. Samyn and J. M. Pearson, Phys. Rev. C \textbf{75}, 064312 (2007).
\bibitem{HFB17} S. Goriely, N. Chamel and J. M. Pearson, Phys. Rev. Lett.
\textbf{102}, 152503 (2009).
\bibitem{Lun03} D. Lunney, J.M. Pearson, C. Thibault, Rev. Mod. Phys.
\textbf{75}, 1021 (2003).

\bibitem{Hivap} Ning Wang, et al., Phys. Rev. C
\textbf{77}, 014603 (2008).

\bibitem{Kowal} M. Kowal, P. Jachimowicz and A. Sobiczewski, Phys. Rev. C
\textbf{82}, 014303 (2010).

\bibitem{Ogan} Yu. Ts. Oganessian et al.,  Phys. Rev. Lett.
\textbf{104}, 142502 (2010).

\bibitem{Dong} Jianmin Dong, Wei Zuo, et al.,
Phys. Rev. C \textbf{81}, 064309 (2010), and references therein.


\bibitem{Ma} Di-Da Zhang, Zhong-Yu Ma, et al., Phys. Rev. C
\textbf{81}, 044319 (2010), and references therein.


\bibitem{Cur} D. Curien, J. Dudek and K. Mazurek, J. Phys.: Conf. Seri.
\textbf{205}, 012034 (2010).

\bibitem{Dudek} J. Dudek, et al., Phys. Rev. Lett. \textbf{88},
252502 (2002).

\end{thebibliography}
\end{document}